\documentclass[12pt]{amsart}
\usepackage{amssymb}
\usepackage{float}
\usepackage{lmodern}
\usepackage{graphicx}
\usepackage[english]{babel}

\theoremstyle{plain}

\numberwithin{equation}{section}

\begin{document}
\title{A generating function and   formulae defining  the first-associated
Meixner-Pollaczek polynomials}
\author{KHALID Ahbli$^{*}$ \ and \ ZOUHA\"{I}R Mouayn$^{\flat}$
}
\maketitle
\begin{center}
\textit{Dedicated to the memory of Professor Ahmed Intissar (1952-2017)
\medskip\\}
\end{center}

\begin{center}
\begin{scriptsize}
${}^{*}$ Department of Mathematics, Faculty of
Sciences, Ibn Zohr University,\vspace{-0.2em}\\ P.O. Box. 8106, Agadir, Morocco.\vspace*{0.2mm}\\
${}^{\flat}$ Department of Mathematics, Faculty of
Sciences and Technics (M'Ghila),\vspace{-0.5em}\\ P.O. Box. 523, B\'{e}ni Mellal, Morocco.
\end{scriptsize}
\end{center}

\thanks{}

\begin{abstract}
While considering nonlinear coherent states with anti-holomorphic coefficients $\bar{z}^{n}/%
\sqrt{x_{n}!}$, we identify  as  first-associated Meixner-Pollaczek polynomials the orthogonal polynomials arising from shift operators which are defined by the sequence $x_{n}=(n+1)^{2}$. We give a formula defining these polynomials
 by writing down their generating function. This also  leads to construct  a  Bargmann-type integral transform whose kernel is given in terms of a $\Psi_1$ Humbert's  function.
\end{abstract}

\section{Introduction}
Coherent states (CSs) were first discovered by Schr\"{o}dinger \cite{schro} as wavepackets having dynamics similar to that of a classical particle submitted to a quadratic potential. They have arised from the study of the quantum harmonic oscillator to become very useful in different areas of physics. CSs  are also present  in the investigation of news nonclassical states of light and their properties. In a such  studies, generalizations of the notion of CSs plays a central role where new concepts such as interferences in phase space and  nonlinear coherent states (NLCSs) have emerged \cite{2}. Furthermore, NLCSs which can be classified as an algebraic generalization of the canonical coherent states of the harmonic oscillator have become an important tool in quantum optics.

In this paper, we replace the factorial $n!$ in  canonical CSs by a generalized
factorial $x_{n}!=x_{1}\cdots x_{n}$ with $x_{0}=0\,\,$and$\,\,x_{n}=(n+1)^{2},\
n=1,2,...$ and we discuss the corresponding resolution of the identity. We proceed by a general method \cite{Siv} to attach to these NLCSs a set of orthogonal polynomials that arise from the shift operators which are defined via the sequence $x_{n}.$ We identify these polynomials as the first-associated Meixner-Pollaczek polynomials which were first introduced in 1950 by Pollaczek  \cite{21} via a tree-terms recurrence relation. A part from this relation no other property seems to be known in the literature. Thus, we establish a formula for their generating function from which we derive their expressions in terms of  Gauss hypergeometric functions $_2F_1$  and classical Meixner-Pollaczek polynomials. Finally, by exploiting  the obtained material we construct a new Bargmann-type integral transform whose kernel is given in terms of a $\Psi_1$ Humbert's confluent hypergeometric function.

The paper is organized as follows. In section 2, we summarize the
construction of NLCSs as well as a procedure to associate to them a set of
orthogonal polynomials. In section 3, we particularize the formalism of NLCSs for the sequence $(n+1)^2$ and we discuss the corresponding resolution of the identity as well as the attached orthogonal polynomials arising from shift operators. For these polynomials a generating function is obtained in section 4. Section 5 is devoted to define a new Bargmann-type transform.

\section{NLCS and polynomials attached to shift operators}

Let $\{x_{n}\}_{n=0}^{\infty },\,\,x_{0}=0$, be an infinite sequence of
positive real numbers. Let $\lim_{n\rightarrow +\infty }x_{n}=R^{2}$, where $%
R>0$ could be finite or infinite, but not zero. We shall use the notation $%
x_{n}!=x_{1}x_{2}\cdots x_{n}$ and $x_{0}!=1$. For each $z\in \mathcal{D}$
some complex domain, a generalized version of canonical CS can be defined
as (\cite{AAG}, p.146) : 
\begin{equation}
|z\rangle =\left(\mathcal{N}(z\bar{z})\right)^{-1/2}\sum\limits_{n=0}^{+\infty }\frac{%
\bar{z}^{n}}{\sqrt{x_{n}!}}|\phi _{n}\rangle ,\quad z\in \mathcal{D}
\label{2.2}
\end{equation}
where the kets $|\phi _{n}\rangle ,\,\,n=0,1,2,...,\infty $, are an
orthonormal basis in an arbitrary (complex, separable, infinite dimensional)
Hilbert space $\mathcal{H}$ and 
\begin{equation}
\mathcal{N}(z\bar{z})=\sum\limits_{n=0}^{+\infty }\frac{(z\bar{z})^{n}}{%
x_{n}!},  \label{2.3}
\end{equation}
is a normalization factor chosen so that the vectors $|z\rangle $ are
normalized to one. These vectors $|z\rangle $ are well defined for all $z$
for which the sum $(\ref{2.3})$ converges, i.e. $\mathcal{D}=\{z\in \mathbb{C%
},|z|<R\}$. We assume that there exists a measure $d\nu $ on $\mathcal{D}$
ensuring the resolution of the identity 
\begin{equation}
\int_{\mathcal{D}}|z\rangle \langle z|d\nu (z,\bar{z})=1_{\mathcal{H}}.
\label{2.4}
\end{equation}
Setting $d\nu (z,\bar{z})=\mathcal{N}(z\bar{z})d\eta (z,\bar{z})$, it is
easily seen that in order for $(\ref{2.4})$ to be satisfied, the measure $%
d\eta $ should be of the form $d\eta (z,\bar{z})=\frac{1 }{2\pi }d\theta
d\lambda (\rho ),\ z=\rho e^{i\theta } \label{2.5} $. The measure $d\lambda $
solves the moment problem 
\begin{equation}
\int_{0}^{R}\rho ^{2n}d\lambda (\rho )=x_{n}!,\quad n=0,1,2,...\ .
\label{2.6}
\end{equation}
In most of practical situations, the support of the measure $d\eta $ is the
whole domain $\mathcal{D}$, i.e., $d\lambda $ is supported on the entire
interval $[0,R)$.

Following \cite{2}, to the above family of coherent states $(\ref{2.2})$ is
naturally associated a set of polynomials $p_n(x)$, orthogonal with respect to some measure on the real line. Moreover, these polynomials may then be used to replace the kets $\left|\phi_n\right\rangle$ in $(\ref{2.2})$. To see this, define the generalized annihilation operator $A$ by its action \textit{"\`a la Glauber"} on the vector $\left| z\right\rangle$ as $A\left| z\right\rangle=z\left|
z\right\rangle$ and its adjoint $A^{\ast}$. Their actions on the basis
vectors $\left| \phi _{n}\right\rangle,\ n=0,1,2,...,$ are easily seen to
be
\begin{eqnarray}  \label{A*}
A\left| \phi _{0}\right\rangle&=&0, \\
A\left| \phi _{n}\right\rangle&=&\sqrt{x_{n}}\left| \phi _{n-1}\right\rangle,
\\
A^{\ast }\left| \phi _{n}\right\rangle&=&\sqrt{x_{n+1}}\left| \phi
_{n+1}\right\rangle.
\end{eqnarray}
Note that $x_n$ are the eigenvalues of the self-adjoint operator $%
H:=A^{\ast}A $, with eigenvectors $\left| \phi _{n}\right\rangle$. We now define the operators, 
\begin{equation}
Q=\frac{1}{\sqrt{2}}(A+A^{\ast}),\ \ \ P=\frac{1}{i\sqrt{2}}(A-A^{\ast}),
\end{equation}
which are analogues of standard position and momentum operators. The
operator $Q$ acts on $\left|\phi_n\right\rangle$ as 
\begin{equation}  \label{2.8}
Q\left|\phi_n\right\rangle=\sqrt{\frac{x_n}{2}}\left|\phi_{n-1}\right\rangle+%
\sqrt{\frac{x_{n+1}}{2}}\left|\phi_{n+1}\right\rangle.
\end{equation}
If now the sum $\sum_{n=0}^{\infty}x_n^{-1/2}$ diverges, the operator $Q$ is
essentially self-adjoint and hence admits a unique self-adjoint extension,
which we again denote by $Q$ (see \cite{AAG}, p.147 and references therein). Let $E_x,\ x\in\mathbb{R}$, be the
spectral family of $Q$, so that, 
\begin{equation}
Q=\int_{\mathbb{R}}xdE_x.
\end{equation}
Thus there is a measure $d\omega(x)=d\left\langle \phi_0|E_x\phi_0\right\rangle$
on $\mathbb{R}$ such that the vectors $\left| \phi _{n}\right\rangle$ can be
realized as elements $p_n(x)$ in $L^2(\mathbb{R}, dw)$. Furthermore, on this
Hilbert space, $Q$ is just the operator of multiplication by $x$ and
consequently, the relation $(\ref{2.8})$ takes the form 
\begin{equation}  \label{6.70}
xp_n(x)=c_np_{n-1}(x)+c_{n+1}p_{n+1}(x), \ \ \ c_n=\sqrt{\frac{x_n}{2}.}
\end{equation}
This is a recursion relation, familiar from the theory of orthogonal
polynomials and thus the functions $p_n(x)$ are polynomials satisfying 
\begin{equation}
\int_{\mathbb{R}}p_n(x)p_m(x)d\omega(x)=\left\langle
\phi_n|\phi_m\right\rangle=\delta_{n,m}.
\end{equation}

\section{NLCS associated with the sequence $x_{n}=(n+1)^{2}$}

According to $(\ref{2.2})$ we associate to the sequence of positive numbers $%
x_{0}=0\,\,\text{and}\,\,x_{n}=(n+1)^{2},\ n=1,2,...$ a set of NLCS by the
following superposition 
\begin{equation}
\left\vert z\right\rangle :=\left( \mathcal{N}(z\bar{z})\right) ^{-\frac{1}{2%
}}\sum_{n=0}^{+\infty }\frac{\bar{z}^{n}}{(n+1)!}\left\vert \phi
_{n}\right\rangle ,  \label{cs}
\end{equation}
of the orthonormal basis vectors $\left\vert \phi _{n}\right\rangle $ in a
Hilbert space $\mathcal{H}$. From the condition 
\begin{equation}
1=\left\langle z|z\right\rangle =\left( \mathcal{N}(z\bar{z})\right)
^{-1}\sum_{n=0}^{+\infty }\frac{(1)_{n}}{(2)_{n}(2)_{n}}\frac{(z\bar{z})^{n}%
}{n!},
\end{equation}%
we see that the normalization factor is given by 
\begin{equation}
\mathcal{N}(z\bar{z})=\frac{1}{(z\bar{z})^{2}}\left( I_{0}(2z\bar{z}%
)-1\right) ,  \label{4.4}
\end{equation}%
$I_{0}$ being the modified Bessel function of the first kind (\cite{23},
p.44). The measure with respect to which these NLCS ensure the resolution of
the identity of the Hilbert space $\mathcal{H}$ as 
\begin{equation}
\int_{\mathbb{C}}\left\vert z\right\rangle \left\langle z\right\vert d\nu
(z)=\mathbf{1}_{\mathcal{H}},  \label{identityresolution}
\end{equation}%
has the form  (\cite{AKM}, p.4): 
\begin{equation}
d\nu (z)=4K_{0}(2|z|)\left( I_{0}(2z\bar{z}
)-1\right) d\mu (z),
\end{equation}
 in terms of the MacDonald function $K_{0}$ (\cite{9},
p.183) and the Lebesgue measure $d\mu $ on $\mathbb{C}$.\medskip\\
\textbf{Remark 3.1.} \textit{Note that in view of} $(\ref{A*})$\textit{ the ket } $|\phi _{n}\rangle $ \textit{can be written as} $|\phi _{n}\rangle =(A^{\ast
})^{n}|\phi _{0}\rangle ,$ \textit{so that the NLCS} $(\ref{cs})$ \textit{may also be obtained by displacing the state} $|\phi _{0}\rangle $ \textit{as} 
\begin{equation}
|z\rangle =\frac{z(A^{\ast })^{-1}}{\sqrt{I_{0}(2z\bar{z})-1}}\left( I_{0}(2%
\bar{z}A^{\ast })-\mathbf{1}\right) |\phi _{0}\rangle ,
\end{equation}%
\textit{where} $\mathbf{1}$ \textit{is the identity operator and} $(A^{\ast })^{-1}$ \textit{is a left
inverse of} $A^{\ast }$ \textit{given by} $(A^{\ast })^{-1}=H^{-1}A$, \textit{where} $H=A^{*}A$.  \textit{Furthermore,
to the sequence} $x_{n}=(n+1)^{2}$ \textit{is attached a function such that} $%
x_{n+1}=f(x_{n})$,\textit{ with} $f(u)=\left( 1+\sqrt{u}\right) ^{2}$. \textit{This function
encodes a generalized Heisenberg algebra (GHA) whose generators} $A$ \textit{and} $%
A^{\ast }$ \textit{satisfy the commutation relation:} $\left[ A,A^{\ast }\right]
=f(H)-H=2\sqrt{H}+1$ (see, \cite{Hassouni}).\newline

Now, we proceed to attach to the sequence $x_{n}=(n+1)^{2}$ a set of orthogonal
polynomials via the three-terms recurrence relation 
\begin{equation}
xp_{n}(x)=\frac{n+2}{\sqrt{2}}p_{n+1}(x)+\frac{n+1}{\sqrt{2}}p_{n-1}(x).
\label{gamma1}
\end{equation}%
\textbf{Proposition 3.1.} \textit{The polynomials satisfying $(\ref{gamma1})$
are the first-associated Meixner-Pollaczek polynomials denoted as} 
\begin{equation}\label{Poly}
p_{n}(x):=P_{n}^{(1/2)}\left( \frac{x}{\sqrt{2}},\frac{\pi }{2},1\right),
\end{equation}%
\textit{and obey the orthonormality relations} 
\begin{equation}
\int_{\mathbb{R}}p_{m}(x)p_{n}(x)\omega (x)dx=\delta _{mn},\quad
m,n=0,1,2,...,
\end{equation}%
\textit{where the weight function} 
\begin{equation}
\omega (x)=\frac{2}{\pi}\left\vert \Gamma \left( \frac{3}{2}+i\frac{x}{\sqrt{2%
}}\right) \right\vert ^{2}\left\vert \ _{2}F_{1}\left( 1,1;\frac{3}{2}+i\frac{x}{\sqrt{2}};\frac{1}{2}\right) \right\vert ^{-2}
\label{weightfun}
\end{equation}\medskip
\textit{is given in terms of the Gauss
hypergeometric sum }$_{2}F_{1}.$

This can be proved by comparing $(\ref{gamma1})$ with the
recurrence relation given in the paper of  Pollaczek (\cite{21}, p.2256) where one can see that polynomials $P_n^{(1/2)}\left(x/\sqrt{2}%
,\pi/2,1\right)$ belong to a larger class of orthogonal polynomials, denoted 
$ P_n^{(\lambda)}(y;\phi, c)$ and called the $c$%
-associated Meixner-Pollaczek polynomials. The latter ones  satisfy 
 
\begin{equation*}
(n+c+1)P_{n+1}^{\lambda}(y;\phi, c)-2\left[(n+\lambda+c)\cos\phi+y\sin\phi%
\right]P_{n}^{\lambda}(y;\phi, c)+(n+2\lambda+c-1)P_{n-1}^{\lambda}(x;\phi,
c)=0,
\end{equation*}
with conditions $0<\varphi<\pi,\ 2\lambda+c>0$ and $c\geq0$ or $0<\phi<\pi,\
2\lambda+c\geq1$ and $c>-1$. For $c=0$, $P_n^{(\lambda)}(y;\phi,0)\equiv
P_n^{(\lambda)}(y;\phi) $ are the well known Meixner-Pollaczek polynomials.\\

Up to our knowledge, a part from the three-terms recurrence relation and orthonormality as written in Pollaczek's paper \cite{21} no other property seems to be known. Here, we first note that the polynomial $P^{(1/2)}_n\left(\frac{x}{\sqrt{2}},\frac{\pi}{2},1\right)$ also admits a representation in terms of the operator $Q$ as follows. From $(\ref{2.8})$, the operator $Q$ can be represented in the $\left\vert
\phi _{n}\right\rangle $ basis as the infinite tridiagonal matrix 
\begin{equation}
Q=\left( 
\begin{array}{cccccc}
0 & \frac{2}{\sqrt{2}} & 0 & 0 & 0 & \cdots \\ 
\frac{2}{\sqrt{2}} & 0 & \frac{3}{\sqrt{2}} & 0 & 0 & \cdots \\ 
0 & \frac{3}{\sqrt{2}} & 0 & \frac{4}{\sqrt{2}} & 0 & \cdots \\ 
0 & 0 & \frac{4}{\sqrt{2}} & 0 & \frac{5}{\sqrt{2}} & \cdots \\ 
0 & 0 & 0 & \frac{5}{\sqrt{2}} & 0 & \cdots \\ 
\vdots & \vdots & \vdots & \vdots & \vdots & \ddots 
\end{array}%
\right) .
\end{equation}%
Let $Q_{n}$ be the truncated matrix consisting of the first $n$ rows and
columns of $Q$. Then it follows that the first-associated Meixner-Pollaczek polynomial is just the characteristic polynomial (up to a scale factor) of $Q_{n}$. That is, 
\begin{equation}
P^{(1/2)}_n\left(\frac{x}{\sqrt{2}},\frac{\pi}{2},1\right)=\frac{2^{\frac{n}{2}}}{(n+1)!}det\left[ xI_{n}-Q_{n}\right] ,
\end{equation}%
where $I_{n}$ is the $n\times n$ identity matrix. Explicitly, 
\begin{equation}
P^{(1/2)}_n\left(\frac{x}{\sqrt{2}},\frac{\pi}{2},1\right)=\frac{2^{\frac{n}{2}}}{(n+1)!}\left\vert 
\begin{array}{cccccccc}
x & -\frac{2}{\sqrt{2}} & 0 & 0 & 0 & \cdots & 0 & 0 \\ 
-\frac{2}{\sqrt{2}} & x & -\frac{3}{\sqrt{2}} & 0 & 0 & \cdots & 0 & 0 \\ 
0 & -\frac{3}{\sqrt{2}} & x & -\frac{4}{\sqrt{2}} & 0 & \cdots & 0 & 0 \\ 
0 & 0 & -\frac{4}{\sqrt{2}} & x & -\frac{5}{\sqrt{2}} & \cdots & 0 & 0 \\ 
0 & 0 & 0 & -\frac{5}{\sqrt{2}} & x & \cdots & \vdots & \vdots \\ 
\vdots & \vdots & \vdots & \vdots & \vdots & \ddots & -\frac{n-1}{\sqrt{2}}
& 0 \\ 
0 & 0 & 0 & 0 & \cdots & -\frac{n-1}{\sqrt{2}} & x & -\frac{n}{\sqrt{2}} \\ 
0 & 0 & 0 & 0 & \cdots & 0 & -\frac{n}{\sqrt{2}} & x 
\end{array}%
\right\vert .
\end{equation}%

The first polynomials of this family are given by{\small {\ 
\begin{eqnarray*}
P^{(1/2)}_0\left(\frac{x}{\sqrt{2}},\frac{\pi}{2},1\right) &=&1, \\
P^{(1/2)}_1\left(\frac{x}{\sqrt{2}},\frac{\pi}{2},1\right) &=&\frac{1}{\sqrt{2}}x, \\
P^{(1/2)}_2\left(\frac{x}{\sqrt{2}},\frac{\pi}{2},1\right) &=&\frac{1}{3}x^{2}-\frac{2}{3}, \\
P^{(1/2)}_3\left(\frac{x}{\sqrt{2}},\frac{\pi}{2},1\right) &=&\frac{\sqrt{2}}{12}x^{3}-\frac{13}{12\sqrt{2}}x, \\
P^{(1/2)}_4\left(\frac{x}{\sqrt{2}},\frac{\pi}{2},1\right) &=&\frac{1}{30}x^{4}-\frac{29}{60}x^{2}+\frac{8}{15}.
\end{eqnarray*}%
}}
 Their graphs are given in Figure 1. \\
\begin{figure}[http]
\centering
\includegraphics[width=0.65\textwidth]{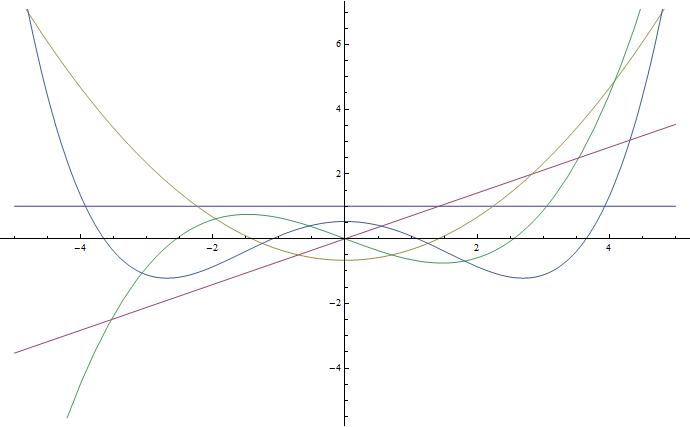}
\caption{{The polynomials $p_{0},p_{1},p_{2},p_{3}$ and $p_{4}$.}}
\end{figure}

\section{A generating function}
In this section, we establish a formula for the generating function for the first-associated Meixner-Pollaczek polynomials, from which we derive an expression of these polynomials in terms of Meixner-Pollaczek polynomials and  Gauss hypergeometric series ${}_2F_1$.\newline
\\
\textbf{Theorem 4.1.} \textit{A generating function for the first
associated Meixner-Pollaczek polynomials} \textit{is given by} 
\begin{eqnarray*}
\qquad\quad\sum\limits_{n\geq 0}P^{(1/2)}_n\left(\frac{x}{\sqrt{2}},\frac{\pi}{2},1\right)t^n=\frac{{}_2F_1\left(1,1;\frac{3+i\sqrt{2}x}{2};\frac{1+it}{2}\right)}{t(i-\sqrt{2}x)} +\frac{{}_2F_1\left(1,1;\frac{3+i\sqrt{2}x}{2};\frac{1}{2}\right)}{\sqrt{2}x-i}\frac{e^{\sqrt{2}x\arctan(t)}}{t\sqrt{t^2+1}},\label{GF new}
\end{eqnarray*}
\textit{for} $|t|<1$ \textit{and} $x\in\mathbb{R}$.\newline
\\
\textbf{Proof.} Let $p_{n}(x):=P_{n}^{(1/2)}\left( x/\sqrt{2},\pi /2,1\right) $ and
denote by $G_{x}(t)$ the generating function $G_{x}(t):=\sum\limits_{n\geq
0}p_{n}(x)t^{n}$. Then, one can check that the function $t\mapsto G_{x}(t)$
satisfies the first order differential equation 
\begin{equation}
(t^{3}+t)G_{x}^{\prime }(t)+(2t^{2}-\sqrt{2}xt+1)G_{x}(t)-1=0
\label{diff equ phi}
\end{equation}%
with the initial condition $G_{x}(0)=1$. Note that polynomials $p_{n}$ are
not monic. However, the renormalized polynomials $q_{n}(x):=c_{n}!p_{n}(x)$
satisfy the recursion relation $q_{n+1}(x)=xq_{n}(x)-c_{n}^{2}q_{n-1}(x)$
from which it is clear that polynomials $q_{n}$ are indeed monic. We first
establish a formula for the generating function of exponential
type $\tilde{G}_{x}(t):=\sum_{n\geq 0}q_{n}(x)t^{n}/n!$, for polynomials $%
q_{n}$ satisfying the recurrence relation 
\begin{equation}
xq_{n}(x)=q_{n+1}(x)+\frac{(n+1)^{2}}{2}q_{n-1}(x).  \label{q_n}
\end{equation}%
By multiplying both sides of Eq.$(\ref{q_n})$ by $t^{n}/n!$ and summing over $%
n $, we see that the function $t\mapsto \tilde{G}_{x}(t)$ has to solve the
second order differential equation 
\begin{equation}
(t^{2}+2)\tilde{G}^{''}_{x}(t)+(5t-2x)\tilde{G}^{'}%
_{x}(t)+4\tilde{G}_{x}(t)=0.  \label{equ.gener}
\end{equation}%
We introduce the following change of variables  $t=i\sqrt{2}%
(1-2\xi)$ and functions  $\varphi (\xi):=\tilde{G}(x,i\sqrt{2}(1-2\xi))$. Then, Eq.$(\ref%
{equ.gener})$ reduces to the hypergeomertic differential equation 
\begin{equation}
\xi(1-\xi)\varphi ^{\prime \prime }(\xi)+\left( \frac{5+i\sqrt{2}x}{2}-5\xi\right)
\varphi ^{\prime }(\xi)-4\varphi (\xi)=0,  \label{hyper-equ}
\end{equation}%
whose solution is of the form (\cite{HBK}, p.256): 
\begin{equation}
\varphi (\xi)=\alpha \ {}_{2}F_{1}\left( 
\begin{array}{c}
2,2 \\ 
\frac{5+i\sqrt{2}x}{2}%
\end{array}%
\big|\xi\right) +\ \beta \xi^{\frac{-3-i\sqrt{2}x}{2}}\ {}_{2}F_{1}\left( 
\begin{array}{c}
\frac{1-i\sqrt{2}x}{2},\frac{1-i\sqrt{2}x}{2} \\ 
\frac{-1-i\sqrt{2}x}{2}%
\end{array}%
\big|\xi\right)\label{4.5}
\end{equation}
where $\alpha ,\beta $ are parameters depending on $x$ and $%
_{2}F_{1}$ denotes the Gauss hypergeometric function. In terms of the
function $t\mapsto \tilde{G}_{x}(t)$, Eq.$(\ref{4.5})$ transforms to 
\begin{equation*}
\tilde{G}_{x}(t)=\alpha (x)\ \ {}_{2}F_{1}\left( 
\begin{array}{c}
2,2 \\ 
\frac{5+i\sqrt{2}x}{2}%
\end{array}%
\big|\frac{2+i\sqrt{2}t}{4}\right) +\beta (x)\left( \frac{2+i\sqrt{2}t}{4}%
\right) ^{\frac{-3-i\sqrt{2}x}{2}}
\end{equation*}%
\begin{equation*}
\times \ {}_{2}F_{1}\left( 
\begin{array}{c}
\frac{1-i\sqrt{2}x}{2},\frac{1-i\sqrt{2}x}{2} \\ 
\frac{-1-i\sqrt{2}x}{2}%
\end{array}%
\big|\frac{2+i\sqrt{2}t}{4}\right) .
\end{equation*}%
To determine $\alpha (x)$ and $\beta (x)$, we first make use of the Euler
transformation (\cite{14}, p.313 ): 
\begin{equation}
{}_{2}F_{1}\left( 
\begin{array}{c}
a,b \\ 
c%
\end{array}%
\big|x\right) =(1-x)^{c-a-b}\ {}_{2}F_{1}\left( 
\begin{array}{c}
c-a,c-b \\ 
c%
\end{array}%
\big|x\right)
\end{equation}%
for $a=\frac{1-i\sqrt{2}x}{2},\ \ b=\frac{1-i\sqrt{2}x}{2}$ and $c=\frac{-1-i%
\sqrt{2}x}{2}$, together with the identity 
\begin{equation}
\left( \frac{1-it}{1+it}\right) ^{\frac{1}{2}iz}=\exp \left( z\arctan
t\right) ,  \label{ident}
\end{equation}%
for $z=\sqrt{2}x$. This gives 
\begin{equation*}
\tilde{G}_{x}(t)=\alpha (x)\ {}_{2}F_{1}\left( 
\begin{array}{c}
2,2 \\ 
\frac{5+i\sqrt{2}x}{2}%
\end{array}%
\big|\frac{2+i\sqrt{2}t}{4}\right) +\beta (x)(2x-t)(2+t^{2})^{-\frac{3}{2}%
}e^{\sqrt{2}x\arctan (\frac{t}{\sqrt{2}})}.
\end{equation*}
Next, the conditions $\tilde{G}_{x}(0)=1$ and $\tilde{G}_{x}^{\prime }(0)=x$ lead to the following system of equations 
\begin{equation}
\left\{ 
\begin{array}{l}
\alpha (x)\ {}_{2}F_{1}\left( 
\begin{array}{c}
2,2 \\ 
\frac{5+i\sqrt{2}x}{2}%
\end{array}%
\big|\frac{1}{2}\right) +\beta (x)(\sqrt{2})^{-1}x=1 \\ 
\alpha (x)\frac{i2\sqrt{2}}{5+i\sqrt{2}x}\ {}_{2}F_{1}\left( 
\begin{array}{c}
3,3 \\ 
\frac{7+i\sqrt{2}x}{2}%
\end{array}%
\big|\frac{1}{2}\right) +\beta (x)(2\sqrt{2})^{-1}(2x^{2}-1)=x.%
\end{array}%
\right.  \label{4.9}
\end{equation}%
The solutions of $(\ref{4.9})$ are then obtained as 
\begin{equation*}
\alpha (x)=-\left[ (2x^{2}-1){}_{2}F_{1}\left( 
\begin{array}{c}
2,2 \\ 
\frac{5+i\sqrt{2}x}{2}%
\end{array}%
\big|\frac{1}{2}\right) -\frac{4\sqrt{2}ix}{5+\sqrt{2}ix}{}_{2}F_{1}\left( 
\begin{array}{c}
3,3 \\ 
\frac{7+i\sqrt{2}x}{2}%
\end{array}%
\big|\frac{1}{2}\right) \right] ^{-1},
\end{equation*}%
and 
\begin{equation*}
\beta (x)=\frac{2^{\frac{3}{2}}x{}_{2}F_{1}\left( 
\begin{array}{c}
2,2 \\ 
\frac{5+i\sqrt{2}x}{2}%
\end{array}%
\big|\frac{1}{2}\right) -\frac{8i}{5+i\sqrt{2}x}{}_{2}F_{1}\left( 
\begin{array}{c}
3,3 \\ 
\frac{7+i\sqrt{2}x}{2}%
\end{array}%
\big|\frac{1}{2}\right) }{(2x^{2}-1){}_{2}F_{1}\left( 
\begin{array}{c}
2,2 \\ 
\frac{5+i\sqrt{2}x}{2}%
\end{array}%
\big|\frac{1}{2}\right) -\frac{4\sqrt{2}ix}{5+\sqrt{2}ix}{}_{2}F_{1}\left( 
\begin{array}{c}
3,3 \\ 
\frac{7+i\sqrt{2}x}{2}%
\end{array}%
\big|\frac{1}{2}\right) }.
\end{equation*}%
To simplify the above expressions, we use the identity (\cite{NIST}, p.388): 
\begin{equation}
z(1-z)(a+1)(b+1)\ {}_{2}F_{1}\left( 
\begin{array}{c}
a+2,b+2 \\ 
c+2%
\end{array}%
\big|z\right)\label{F}
\end{equation}%
\begin{equation*}
+\newline
(c-(a+b+1)z)(c+1)\ {}_{2}F_{1}\left( 
\begin{array}{c}
a+1,b+1 \\ 
c+1%
\end{array}
\big|z\right) -c(c+1)\ {}_{2}F_{1}\left( 
\begin{array}{c}
a,b \\ 
c
\end{array}
\big|z\right) =0
\end{equation*}
for $a=b=1,\ c=\frac{3+i\sqrt{2}x}{2},\ z=\frac{1}{2}$ and $a=b=0,\ c=\frac{%
1+i\sqrt{2}x}{2},z=\frac{1}{2}$. Therefore, 
\begin{equation}
\alpha (x) =\frac{1}{(1+i\sqrt{2}x)(3+i\sqrt{2}x)}\quad,\quad \beta (x) =\frac{2i}{1+i\sqrt{2}x}{}_{2}F_{1}(1,1;\frac{3+i\sqrt{2}x}{2};%
\frac{1}{2}).
\end{equation}%
Summarizing the above calculations, we obtain that 
\begin{equation}\label{eq4.11}
\tilde{G}_{x}(t)=\left( (1+i\sqrt{2}x)(3+i\sqrt{2}x)\right)
^{-1}{}_{2}F_{1}\left( 
\begin{array}{c}
2,2 \\ 
\frac{5+i\sqrt{2}x}{2}%
\end{array}%
\big|\frac{1}{2}+i\frac{t}{2\sqrt{2}}\right)
\end{equation}
\begin{equation*}
+{}_{2}F_{1}\left( 
\begin{array}{c}
1,1 \\ 
\frac{3+i\sqrt{2}x}{2}%
\end{array}%
\big|\frac{1}{2}\right) \frac{(4x-2t)e^{\sqrt{2}x\arctan (\frac{t}{\sqrt{2}}%
)}}{(\sqrt{2}x-i)(t^{2}+2)^{\frac{3}{2}}}
\end{equation*}%
for every $x\in \mathbb{R}$. Using $(\ref{eq4.11})$ and the relation 
\begin{equation}
G_{x}(t)=\frac{t^{2}+1}{\sqrt{2}xt-t^{2}}\tilde{G}_{x}(\sqrt{2}t)+\frac{1}{%
t^{2}-\sqrt{2}xt}  \label{G_phii}
\end{equation}%
 connecting the two generating functions, we get
\begin{eqnarray}  \label{generatingfunction}
\sum\limits_{n\geq 0}p_n(x)t^n=-\frac{1}{t(\sqrt{2}x-t)}+\frac{(t^2+1){}_2F_1\left(2,2;\frac{%
5+i\sqrt{2}x}{2};\frac{1+it}{2}\right)}{t(\sqrt{2}x-t)(1+i\sqrt{2}x)(3+i%
\sqrt{2}x)}
\end{eqnarray}
\begin{equation*}
+\frac{{}_2F_1\left(1,1;\frac{3+i\sqrt{2}x}{2};\frac{1}{2}\right)}{\sqrt{2}%
x-i}\frac{e^{\sqrt{2}x\arctan(t)}}{t\sqrt{t^2+1}} .
\end{equation*}
Finally, we arrive at the expression $(\ref{GF new})$ by using the following equality 
\begin{eqnarray*}
\frac{(t^2+1){}_2F_1\left(2,2;\frac{5+i\sqrt{2}x}{2};\frac{1+it}{2}\right)}{t(\sqrt{2}x-t)(1+i\sqrt{2}x)(3+i\sqrt{2}x)}=\frac{1}{t(\sqrt{2}x-t)}+\frac{{}_2F_1\left(1,1;\frac{3+i\sqrt{2}x}{2};\frac{1+it}{2}\right)}{t(i-\sqrt{2}x)} 
\end{eqnarray*}
which can be derived from $(\ref{F})$ by an appropriate choice of parameters.$\qquad\qquad\square$\medskip\\
\textbf{Corollary 4.1. }\textit{The following identity} 
\begin{equation}
{}_{2}F_{1}\left( 
\begin{array}{c}
2,2 \\ 
\frac{5}{2}%
\end{array}%
\big|\frac{1+it}{2}\right) =\frac{3}{t^{2}+1}\left[ \frac{t}{\sqrt{t^{2}+1}}%
\left( \frac{i\pi }{2}-Log\left( t+\sqrt{t^{2}+1}\right) \right) +1\right]
\label{Hyp}
\end{equation}%
\textit{holds true for }$|t|<1$.\medskip\\
\textbf{Proof.} Evaluating at $x=0$ the closed form of the generating function as given in the
right hand side of $(\ref{generatingfunction})$, leads to
\begin{equation}
\sum_{n\geq 0}P_{n}^{(1/2)}\left( 0,\frac{\pi }{2},1\right) t^{n}=\frac{2%
\sqrt{t^{2}+1}-it\pi }{2t^{2}\sqrt{t^{2}+1}}-\frac{t^{2}+1}{3t^{2}}{}%
_{2}F_{1}\left( 
\begin{array}{c}
2,2 \\ 
\frac{5}{2}%
\end{array}%
\big|\frac{1+it}{2}\right) .  \label{4.18}
\end{equation}%
By another side, we use the recurrence relation $(\ref{gamma1})$ to get
the following evaluations of polynomials $P_{n}^{(1/2)}\left( x,\frac{\pi }{2%
},1\right) $ at $x=0$ as 
\begin{equation}
\left\{ 
\begin{array}{lll}
P^{(1/2)}_{2n}\left(0,\frac{\pi}{2},1\right)=(-1)^{n}\frac{2^{2n}(n!)^{2}}{(2n+1)!}, \\ 
P^{(1/2)}_{2n+1}\left(0,\frac{\pi}{2},1\right)=0. 
\end{array}
\right.
\end{equation}%
This allows us to express the left hand side of $(\ref{generatingfunction})$
with $x=0$ as 
\begin{equation}
G_{0}(t)=\frac{1}{2t}\sum\limits_{n\geq 0}(-1)^{n}\frac{(n!)^{2}}{(2n+1)!}%
(2t)^{2n+1}.  \label{g_0}
\end{equation}%
To the later one, we apply the formula (\cite{prud1}, p.714): 
\begin{equation}
\sum\limits_{n\geq 0}(-1)^{n}\frac{(n!)^{2}}{(2n+1)!}u^{2n+1}=4(4+u^{2})^{-%
\frac{1}{2}}Log\left( \frac{u}{2}+\sqrt{\frac{u^{2}}{4}+1}\right) ,\ \ \
|u|<2,
\end{equation}%
for $u=2t$, to obtain an expression for $G_{0}(t)$ as
\begin{equation}
G_{0}(t)=\frac{Log\left( t+\sqrt{t^{2}+1}\right) }{t\sqrt{t^{2}+1}},\qquad
|t|<1.  \label{g02}
\end{equation}%
By equating $(\ref{4.18})$ with $(\ref{g02})$, we arrive at the identity $(\ref{Hyp})$.        $\square$
\\
\\
\textbf{Remark 4.1.} Using the fact that $i \arcsin(z)=Log(iz+%
\sqrt{1-z^2})$ in $(\ref{Hyp})$ and taking $t=-i(2 \xi -1)$, we recover the identity (\cite{25}, p.481): 
\begin{equation}
{}_2F_1\left(%
\begin{array}{c}
2,2 \\ 
\frac{5}{2}%
\end{array}
\big|\xi\right)=\frac{3}{4\xi}\left(1-\xi\right)^{-1}\left[1-\frac{1-2\xi}{%
\sqrt{\xi(1-\xi)}}\arcsin\left(\sqrt{\xi}\right)\right].\vspace*{5mm}
\end{equation}
\textbf{Theorem 4.2.} \textit{The first-associated Meixner-Pollaczek polynomials can be written as}
\begin{eqnarray*}\label{P_n}
P^{(1/2)}_n\left(x,\frac{\pi}{2},1\right)=\frac{i^{n}(2)_{n}\,{}_2F_1\left(n+2,n+2;n+\frac{5}{2}+ix;\frac{1}{2}\right)}{2^{n+2}\left(\frac{1}{2}+ix\right)_{n+2}} +\frac{{}_2F_1\left(1,1;\frac{3}{2}+ix;\frac{1}{2}\right)}{(2x-i)}P_{n+1}^{(1/2)}\left( x,\frac{\pi }{2}
\right).
\end{eqnarray*}
\textit{in terms of $_2F_1$-sums and Meixner-Pollaczek polynomials}.\medskip\medskip\\
\textbf{Proof.} By multiplying by the variable $t$ the generating function $G_x(t)$ in $(\ref{GF new})$ as 
\begin{eqnarray*}
tG_x(t)=\frac{{}_2F_1\left(1,1;\frac{3+i\sqrt{2}x}{2};\frac{1+it}{2}\right)}{(i-\sqrt{2}x)} +\frac{{}_2F_1\left(1,1;\frac{3+i\sqrt{2}x}{2};\frac{1}{2}\right)}{\sqrt{2}x-i}\frac{e^{\sqrt{2}x\arctan(t)}}{\sqrt{t^2+1}}
\end{eqnarray*}
and denoting
\begin{equation}
g_x(t):=\frac{e^{\sqrt{2}x\arctan(t)}}{\sqrt{t^2+1}}=\sum_{n=0}^{+\infty }P_{n}^{(1/2)}\left( \frac{x}{\sqrt{2}},\frac{\pi }{2}%
\right) t^{n}, 
\end{equation}
the $(n+1)$ derivative of the function $t\longmapsto tG_x(t)$ reads
\begin{eqnarray}\label{DG1}
\frac{\partial^{n+1}}{\partial t^{n+1}}\left( tG_x(t)\right)=\frac{i^{n}(2)_{n}(2)_{n}\,{}_2F_1\left(n+2,n+2;n+\frac{5+i\sqrt{2}x}{2};\frac{1+it}{2}\right)}{2^{n+1}\left(\frac{3+i\sqrt{2}x}{2}\right)_{n+1}(1+i\sqrt{2}x)}\\ +\frac{{}_2F_1\left(1,1;\frac{3+i\sqrt{2}x}{2};\frac{1}{2}\right)}{\sqrt{2}x-i}\frac{\partial^{n+1}}{\partial t^{n+1}}g_x(t).\nonumber
\end{eqnarray}
By another hand, one has
\begin{eqnarray}\label{DG2}
\frac{\partial^{n+1}}{\partial t^{n+1}}(tG_x(t))=(n+1)\frac{\partial^{n}}{\partial t^{n}}G_x(t)+t\frac{\partial^{n+1}}{\partial t^{n+1}}G_x(t).
\end{eqnarray}
From $(\ref{DG1})$-$(\ref{DG2})$, it then follows that 
\begin{eqnarray}\label{DG1=DG2}
&&(n+1)\frac{\partial^{n}}{\partial t^{n}}G_x(t)+t\frac{\partial^{n+1}}{\partial t^{n+1}}G_x(t)=\\ &&\frac{i^{n}(2)_{n}(2)_{n}\,{}_2F_1\left(n+2,n+2;n+\frac{5+i\sqrt{2}x}{2};\frac{1+it}{2}\right)}{2^{n+1}\left(\frac{3+i\sqrt{2}x}{2}\right)_{n+1}(1+i\sqrt{2}x)} +\frac{{}_2F_1\left(1,1;\frac{3+i\sqrt{2}x}{2};\frac{1}{2}\right)}{\sqrt{2}x-i}\frac{\partial^{n+1}}{\partial t^{n+1}}g_x(t).\nonumber
\end{eqnarray}
Next, by evaluating the last equation at $t=0$ and using the fact that
\begin{equation}
P^{(1/2)}_n\left(\frac{x}{\sqrt{2}},\frac{\pi}{2},1\right)=\frac{1}{n!}\frac{\partial^n}{\partial t^n}G_x(t)\mid_{t=0},
\end{equation}
we arrive at
\begin{eqnarray*}
P^{(1/2)}_n\left(\frac{x}{\sqrt{2}},\frac{\pi}{2},1\right)=\frac{i^{n}(2)_{n}\,{}_2F_1\left(n+2,n+2;n+\frac{5+i\sqrt{2}x}{2};\frac{1}{2}\right)}{2^{n+2}\left(\frac{1+i\sqrt{2}x}{2}\right)_{n+2}} +\frac{{}_2F_1\left(1,1;\frac{3+i\sqrt{2}x}{2};\frac{1}{2}\right)}{(\sqrt{2}x-i)}P_{n+1}^{(1/2)}\left( \frac{x}{\sqrt{2}},\frac{\pi }{2}%
\right).
\end{eqnarray*}
Finally, changing $x$ by $\sqrt{2}x$ completes the proof.     $\square$
\section{A Bargmann-type integral transform}

Following (\cite{2}, p.4) the  orthogonal polynomials $%
 p_{n}(x)$ arising from the shift operators that are defined by the sequence $x_n$ may then be used to replace the abstract  ket vectors $%
\left\vert \phi _{n}\right\rangle $ in $(\ref{2.2})$. In our case,  this  means that the
obtained first-associated Meixner-Pollaczek polynomials $\left( \ref{Poly}\right) $
may define eigenstates of some explicit Hamiltonian operator. Such an operator could be determined by
applying the method in \cite{borzov} for example. In a such  way,  the wave functions of the
resulting NLCSs   as vectors in $L^2(\mathbb{R},\omega(x) dx)$, the Hilbert space spanned by the first-associated Meixner-Pollaczek polynomials, are of the form
\begin{equation}
\langle \xi|z\rangle :=\left( \mathcal{N}(z\bar{z})\right) ^{-1/2}\sum\limits_{n\geq 0}%
\frac{\overline{z}^{n}}{\sqrt{x_{n}!}}P_{n}^{(1/2)}\left( \frac{\xi}{\sqrt{2}},%
\frac{\pi }{2},1\right),\;\xi\in\mathbb{R}.   \label{5.1}
\end{equation}
\textbf{Proposition 5.1. }
\textit{The coordinate space representation of the NLCSs (5.1) is given by}
 
\begin{equation}\label{5.2}
\langle \xi|z\rangle =\frac{ze^{-i\bar{z}}}{\sqrt{I_0(2z\bar{z})-1}(\sqrt{2}\xi-i)}\big[ {}_2F_1\left(1,1;\frac{3+i\sqrt{2}\xi}{2};\frac{1}{2}\right) {}_1F_1\left(\frac{1}{2}-\frac{i\xi}{\sqrt{2}},1;2i\bar{z}\right)
\end{equation}
\begin{equation*}
-\Psi_1\left(1,1;\frac{3+i\sqrt{2}\xi}{2},1;\frac{1}{2},i\bar{z}\right)\big]
\end{equation*} 
\textit{for every} $\xi \in\mathbb{R}$, \textit{in terms of the Humbert's $\Psi_1$-series}. 
\medskip\\
\textbf{Proof.} By using Theorem $4.2$, we replace polynomials in (5.1) by their explicit expressions. This gives  
\begin{eqnarray}
\mathcal{S}:&=&\sum_{n\geq 0}P_{n}^{(1/2)}\left( \frac{\xi}{\sqrt{2}},\frac{\pi }{2},1\right)\frac{\bar{z}^n}{(n+1)!}\label{5.3}\\
&=&\frac{1}{2(1+i\sqrt{2}\xi)}\sum_{n\geq 0}\frac{{}_2F_1\left(n+2,n+2;n+\frac{5+i\sqrt{2}\xi}{2};\frac{1}{2}\right)}{2^{n}\left(\frac{3+i\sqrt{2}\xi}{2}\right)_{n+1}} (i\bar{z})^n \nonumber\\
&+&\frac{{}_2F_1\left(1,1;\frac{3+i\sqrt{2}\xi}{2};\frac{1}{2}\right)}{(\sqrt{2}\xi-i)}\sum_{n\geq 0}P_{n+1}^{(1/2)}\left( \frac{\xi}{\sqrt{2}},\frac{\pi }{2}%
\right)\frac{\bar{z}^n}{(n+1)!}\nonumber
\end{eqnarray}
where the last sum in the R.H.S of the last equation involves a generating function for the classical Meixner-Pollaczek polynomials as follows 
\begin{eqnarray}
\sum_{n\geq 0}P_{n+1}^{(1/2)}\left( \frac{\xi}{\sqrt{2}},\frac{\pi }{2}%
\right)\frac{\bar{z}^n}{(n+1)!}=\frac{1}{\bar{z}} e^{-i\bar{z}} {}_1F_1\left(\frac{1}{2}-\frac{i\xi}{\sqrt{2}},1;2i\bar{z}\right)-\frac{1}{\bar{z}}.
\end{eqnarray}
Therefore, the sum $\mathcal{S}$ in $(\ref{5.3})$ takes the form 
\begin{equation*}
\mathcal{S}=\frac{1}{\bar{z}(\sqrt{2}\xi-i)}\big[ {}_2F_1\left(1,1;\frac{3+i\sqrt{2}\xi}{2};\frac{1}{2}\right)e^{-i\bar{z}} {}_1F_1\left(\frac{1}{2}-\frac{i\xi}{\sqrt{2}},1;2i\bar{z}\right)
\end{equation*}
\begin{equation}\label{5.4}
-\sum_{n\geq 0}\frac{1}{\left(\frac{3+i\sqrt{2}\xi}{2}\right)_{n}}\left(\frac{i\bar{z}}{2} \right)^n{}_2F_1\left(n+1,n+1;n+\frac{3+i\sqrt{2}\xi}{2};\frac{1}{2}\right)\big].
\end{equation}
For the remaining series in $(\ref{5.4})$, we prove that (see Appendix A):
\begin{eqnarray*}
\sum_{n\geq 0}\frac{1}{\left(\frac{3+i\sqrt{2}\xi}{2}\right)_{n}}\left(\frac{i\bar{z}}{2} \right)^n{}_2F_1\left(n+1,n+1;n+\frac{3+i\sqrt{2}\xi}{2};\frac{1}{2}\right)=e^{-i\bar{z}}\Psi_1\left(1,1;\frac{3+i\sqrt{2}\xi}{2},1;\frac{1}{2},i\bar{z}\right),
\end{eqnarray*} 
where $\Psi_1$ is a Humbert's confluent hypergeometric function of two variables defined by (\cite{AKF}, p.126): 
\begin{equation}
\Psi_1(\alpha,\beta;\gamma,\gamma ';x,y)=\sum_{m,n\geq 0}\frac{(\alpha)_{m+n}(\beta)_m}{(\gamma)_m(\gamma ')_n}\frac{x^m}{m!}\frac{y^n}{n!}, \qquad |x|<1.
\end{equation}
Finally, recalling the expression of the  prefactor $\mathcal{N}(z)$ in $(\ref{2.3})$ and summarizing the above calculations, we arrive at the announced result $(\ref{5.2})$. $\qquad\qquad\qquad\qquad\qquad\qquad\square$
\medskip\\
\textbf{Corollary 5.1. } 
\textit{In addition, the function} 
\begin{equation}
 \Lambda(\xi,z):= \frac{\sqrt{I_0(2z\bar{z})-1}}{z\bar{z}}\langle \xi|\bar{z}\rangle =\displaystyle\sum_{n\geq 0}\frac{z^n}{\sqrt{x_n!}}P_{n}^{(1/2)}\left( \frac{\xi}{\sqrt{2}},\frac{\pi }{2},1\right),
\end{equation}
\textit{is also a generating function for  the first-associated Meixner-Pollaczek  polynomials in the sense that }
\begin{equation}
P_{n}^{(1/2)}\left( \frac{\xi}{\sqrt{2}},\frac{\pi}{2},1\right)=(n+1)\frac{\partial^{n}}{\partial z^n}\Lambda(\xi,z)|_{z=0}.
\end{equation}

\medskip

Once we have obtained a closed form for the NLCSs (5.1) we can define the associated coherent states transform. The latter one should map the Hilbert space $L^2(\mathbb{R},\; \omega(x)dx)$ with\\
\begin{equation}
\omega (x)=\frac{2}{\pi}\left\vert \Gamma \left( \frac{3}{2}+i\frac{x}{\sqrt{2%
}}\right) \right\vert ^{2}\left\vert \ _{2}F_{1}\left( 1,1;\frac{3}{2}+i\frac{x}{\sqrt{2}};\frac{1}{2}\right) \right\vert ^{-2},
\end{equation}
onto the Hilbert space $\mathcal{A}(\mathbb{C})$ of complex-valued analytic functions on $\mathbb{C}$, which are square integrable with respect to the measure  $d\nu(z)=4(z\bar{z}%
)^{2}K_{0}(2|z|)d\mu (z)$. The following Theorem makes this statement more precise. \medskip\\
\\
\textbf{Theorem 5.1. } \textit{The NLCSs (5.1) give rise to a  Bargmann-type  transform  through the unitary embedding } $\mathcal{B}%
:L^{2}(\mathbb{R},\omega (\xi)d\xi)\rightarrow \mathcal{A}(\mathbb{C})\subset L^2(\mathbb{C},d\nu)$ \textit{defined by} 
\begin{equation}
 \mathcal{B}[\varphi ](z)=\int_{\mathbb{R}}\mathcal{B}(z,\xi)\varphi(\xi)\omega(\xi)d\xi,\label{BAR}
\end{equation}%
\textit{where } 
$$\mathcal{B}(z,x)=\frac{e^{-iz}}{z(\sqrt{2}\xi-i)}\big[ {}_2F_1\left(1,1;\frac{3+i\sqrt{2}\xi}{2};\frac{1}{2}\right) {}_1F_1\left(\frac{1}{2}-\frac{i\xi}{\sqrt{2}},1;2iz\right)
\\
-\Psi_1\left(1,1;\frac{3+i\sqrt{2}\xi}{2},1;\frac{1}{2},iz\right)\big].\medskip$$

With the help of this transform, we see that any arbitrary state $|\varphi\rangle$ in $L^{2}(\mathbb{R},\omega (\xi)d\xi)$ has a representation in terms of the NLCSs (5.1) as follows
\begin{equation}
|\varphi\rangle =4\int_{\mathbb{C}} d\mu(z)z\bar{z}\sqrt{I_0(2z\bar{z})-1}K_0(2|z|) \bar{\mathcal{B}[\varphi](z)}|z\rangle ,
\end{equation}
in terms of the Lebesgue measure $d\mu(z)$ on $\mathbb{C}$. Therefore, the norm square  also reads
\begin{equation}
\langle\varphi|\varphi\rangle = 4\int_{\mathbb{C}} d\mu(z)(z\bar{z})^2K_0(2|z|) \left|\mathcal{B}[\varphi ](z)\right|^2
\end{equation}  
for every $|\varphi\rangle$ in $L^{2}(\mathbb{R},\omega (\xi)d\xi)$.
\bigskip 
\begin{center}
{\large\textbf{Appendix A}} \medskip\\
\end{center}

To obtain a closed form for the series 
\begin{equation}
\delta(t,c;x):= \sum_{n\geq 0}\frac{t^n}{\left(c\right)_{n}}{}_2F_1\left(n+1,n+1;n+c;x\right), \tag{A.1}
\end{equation}
 we make use the integral representation of the $_2F_1$-sum  (\cite{25}, p.431):
\begin{equation}
   {}_2F_1(\alpha, \beta; \gamma; z)=\frac{\Gamma(\gamma)}{\Gamma(\alpha)\Gamma(\gamma -\alpha)}\int_{0}^{1}t^{\alpha-1}(1-zt)^{-\beta}(1-t)^{\gamma-\alpha-1}dt, \tag{A.2}
   \end{equation}
$Re \gamma >Re \alpha >0$ and $|\arg(1-z)|<\pi$, for $\alpha=\beta=n+1$, $\gamma=n+c$ and $z=x$. Thus,
\begin{equation}\label{A.3}
\delta(t,c;x)= (c-1)\int_{0}^{1}(1-s)^{c-2}(1-xs)^{-1}exp(\frac{st}{1-xs})ds.\tag{A.3}
\end{equation}
Next, by using the generating function for Laguerre polynomials (\cite{askey}, p.242):
\begin{equation*}
(1-u)^{-\alpha-1}exp(\frac{yu}{u-1})=\sum_{j\geq 0}u^j L_j^{\alpha}(y),
\end{equation*}
for $u=xs$, $y=-\frac{t}{x}$ and $\alpha=0$, the R.H.S of $(\ref{A.3})$ can be written as 
\begin{equation}\label{A.4}
\delta(t,c;x)=  (c-1)\sum_{j\geq 0}L_j\left(-\frac{t}{x}\right)x^j\int_{0}^{1}(1-s)^{c-2}s^jds. \tag{A.4}
\end{equation}
writing
\begin{equation*}
\frac{(1)_j}{(c)_j}=(c-1)\int_{0}^{1}(1-s)^{c-2} s^{j}ds
\end{equation*}
then  $(\ref{A.4})$ reduces to 
\begin{equation}
\delta(t,c;x) =\sum_{j\geq 0}\frac{(1)_j}{(c)_j}L_j\left(-\frac{t}{x}\right)x^j. \tag{A.5}
\end{equation}
We are now in position to exploit the formula (\cite{S}, p.152): 
 \begin{equation*}
\sum_{j\geq 0}\frac{(\lambda)_j}{(\mu)_j}L_j^{(\alpha)}\left(y\right)\zeta^j=e^{y}\Psi_1(\alpha+1,\lambda;\mu,\alpha+1;\zeta,-y) \tag{A.6}
\end{equation*}
for parameters $\lambda=1,\;\mu=c,\;\alpha=0,\;y=\frac{-t}{x}$ and $\zeta=x$ where  $\Psi_1$ is a Humbert's confluent hypergeometric function of two variables defined by (\cite{AKF}, p.126) :
\begin{equation*}
\Psi_1(\alpha,\beta;\gamma,\gamma';x,y)=\sum_{m,n\geq 0}\frac{(\alpha)_{m+n}(\beta)_m}{(\gamma)_m(\gamma')_n}\frac{x^m}{m!}\frac{y^n}{n!}.\tag{A.7}
\end{equation*} 
One get
\begin{equation}
\sum_{n\geq 0}\frac{t^n}{\left(c\right)_{n}}{}_2F_1\left(n+1,n+1;n+c;x\right)
=e^{-\frac{t}{x}}\Psi_1\left(1,1;c,1;x,\frac{t}{x}\right). \tag{A.8}
\end{equation}

Finally, by replacing $t$ by $\frac{i\bar{z}}{2}$
, $c$ by $(3+i\sqrt{2}\xi)/2$ and $x$ by $1/2$, the proof of (5.5) is completed. $\square$

\end{document}